\documentclass[aps,preprint,letterpaper,amsmath,amssymb]{revtex4}
\usepackage{graphicx}
\usepackage{color}
\begin{document}
\newcommand{\rf}[1]{(\ref{#1})}

\newcommand{\bfomega}{ \mbox{\boldmath{$\omega$}}}

\title{Dynamic particle tracking reveals
the  aging temperature of a colloidal glass}
%

\author{ Ping Wang, Chaoming Song, Hern\'an A. Makse }
\affiliation{ Levich Institute and Physics Department, City
College of New York, New York, NY 10031 }


\begin{abstract}
Understanding glasses is considered to be one of the most
fundamental problems in statistical physics. A theoretical
approach to unravel their universal properties is to consider the
validity of equilibrium concepts such as temperature and
thermalization in these out-of-equilibrium systems. Here we
investigate the autocorrelation and response function to monitor
the aging of a colloidal glass. At equilibrium, all the
observables are stationary while in the out-of-equilibrium glassy
state they have an explicit dependence on the age of the system.
We find that the transport coefficients scale with the aging-time
as a power-law, a signature of the slow relaxation. Nevertheless,
our analysis reveals that the glassy system has thermalized at a
{\it constant} temperature independent of the age and larger than
the bath, reflecting the structural rearrangements of
cage-dynamics. Furthermore, a universal scaling law is found to
describe the global and local fluctuations of the observables.



\end{abstract}

\maketitle

\newpage

{\bf Introduction.---} Increasing the volume fraction of a
colloidal system slows down the Brownian dynamics of its
constitutive particles, implying a limiting density, $\phi_g$,
above which the system can no longer be equilibrated with its bath
\cite{Pusey}. Hence, the thermal system falls out of equilibrium
on the time scale of the experiment and thus undergoes a {\it
glass transition} \cite{Cugliandolo}. Even above $\phi_g$ the
particles continue to relax, but the nature of the relaxation is
very different to that in equilibrium. This phenomenon of a
structural slow evolution beyond the glassy state is known as
``aging'' \cite{Struik}.  The system is no longer stationary and
the relaxation time is found to increase with the age of the
system, $t_w$, as measured from the time of sample preparation.

This picture applies not only to structural glasses such as
colloids, silica and polymer melts, but also to spin-glasses,
ferromagnetic coarsening, elastic manifolds in quenched disorder
and jammed matter such as grains and emulsions
\cite{Cugliandolo,Kurchan_prl,Bouchaud}. Theories originally
developed in the field of spin-glasses \cite{Kirkpatrick} attempt
to develop
a common framework for the understanding of aging. For example,
the structural glass and spin-glass transitions have been coupled
by the low temperature extension of the Mode-Coupling Theory (MCT)
\cite{Kurchan_MCT,Kurchan_prl,Bouchaud}.
More generally, this approach is related to analogous ideas
developed in the field of granular matter such as compactivity
\cite{Edwards,Ono,Makse_nature,Anna,Song}, and the inherent
structure formalisms \cite{Stillinger}, adapted to the energy
landscape of glasses \cite{Sciortino}.

One of the important features of this scenario is a separation of
time-scales where the observables are equilibrated at different
temperatures,
even though the system is  far from equilibrium. While theoretical
results have flourished, the difficulties in the experimental
testing of the fundamental predictions of the theories have
hampered the development of an understanding of aging in glasses
\cite{Israeloff,Bellon,Herisson,Buisson,Abou}.
Experiments so far have shown conflicting results which are
usually masked by large intermittent fluctuations in the
observables \cite{Buisson}; a behavior which seems beyond the
current theoretical formalisms \cite{Cugliandolo}.
On the other hand, some numerical results are more favorable
\cite{Parisi,Barrat_Kob,Berthier}. Furthermore, the concept of
temperature has been shown to be useful to describe other far form
equilibrium systems such as non-thermal granular materials
\cite{Ono,Makse_nature,Anna,Song}.

Here we use a model glass which
is one of the simplest systems undergoing a glass transition: a
colloidal glass of micrometer size particles, where the
interactions between particles can be approximated as hard core
potentials \cite{Pusey,Kegel,Weeks_sci}. The system is index
matched to allow the visualization of tracer particles in the
microscope \cite{Weeks_sci}.
Owing to the simplicity of the system, we are able to follow the
trajectories of magnetic tracers embedded in the colloidal sample
and use this information as an ideal ``thermometer" to measure the
temperature for the different modes of relaxation. In turn, we
measure the autocorrelation function of the displacements and the
integrated response to an external magnetic field as an indicator
of the dynamics via a Fluctuation-Dissipation Theorem (FDT).
For this system we show that, even though the diffusivities and
mobilities of the tracers scale with the age of the system, there
exists an effective temperature which is uniquely defined and
remains constant independent of the age. This effective
temperature is larger than the bath temperature and controls the
slow relaxation of the system, as if the system were at
``equilibrium". We find a scaling behavior with the waiting time,
which describes in a unified way not only the global but also the
local fluctuations of the correlations and responses as well as
the cage dynamics in the system.

{\bf Correlations and responses.---} Our experiments use a
colloidal suspension consisting of a mixture of
poly-(methylmethacrylate) (PMMA) sterically stabilized colloidal
particles (radius $a_{p}=1.5\ \mu m$, density $\rho_{p}=1.19\
g/cm^{3}$, polydispersity $\sim 14\%$) plus a small fraction of
superparamagnetic beads (radius $a_{m}=1.6\ \mu m$ and density
$\rho_{m}=1.3\ g/cm^{3}$, from Dynal Biotech Inc.) as the tracers
(More details in the {\bf Method} section). In order to
investigate the dynamical properties of the aging regime, we first
consider the autocorrelation function as the mean square
displacement (MSD) averaged over 82 tracer particles, $C(t,t_w)
\equiv \langle\Delta x^{2}(t, t_w)\rangle/2=\langle[x(t_w+\Delta
t, t_w)-x(t_w, t_w)]^2\rangle/2$, at a given observation time,
$t=t_w+\Delta t$, after the sample has been aging for $t_w$ as
measured from the end of the stirring process (see Appendix
\ref{data}). Then, we measure the integrated response function (by
adding the external magnetic force, $F$) given by the average
position of the tracers, $\chi(t,t_w) \equiv \langle x(t_w +\Delta
t, t_w ) - x(t_w ,t_w ) \rangle /F$.

Analytical extensions of the MCT for supercooled liquids to the
low temperature regime of glasses allow for the interpretation  of
the aging of global correlation and response functions
\cite{Cugliandolo,Bouchaud}.
In these frameworks, the evolution of $C(t,t_w)$ and $\chi(t,t_w)$
are separated into a stationary part (short time) and an aging
part (long time): $C(t,t_w)=C_{\rm st}(t-t_w)+C_{\rm ag}(t,t_w)$
and $\chi(t,t_w)=\chi_{\rm st}(t-t_w)+\chi_{\rm ag}(t,t_w)$, where
we have included the explicit dependence on $t_w$ in the aging
part.

This result can be rationalized in terms of the so-called ``cage
dynamics". As the density of the system increases the particles
are trapped in cages.  The motion inside the cage is still
equilibrated at the bath temperature and is determined by the
Gibbs distribution of states. This dynamics gives rise to the
stationary part of the response and correlation functions, which
satisfy the usual equilibrium relations such as the FDT.  However,
the correlation does not decay to zero but remains constant since
particles are trapped in cages for a long time.  Thermal activated
motions lead to a second structural relaxation which is
responsible for the aging part of the dynamics. In this regime the
system is off equilibrium and correlations and responses depend
not only on the time of observation $t$ but also on the waiting
time, $t_w$. Slow, non-exponential relaxation ensues and a strong
violation of the FDT is expected. However, the interesting result
is that this breakdown leads to a new definition of temperature
for the slow modes which has been proposed to be the starting
point of a unifying description of aging in glassy systems
\cite{Cugliandolo}.

\begin{figure}
\centering \resizebox{15cm}{!}{\includegraphics{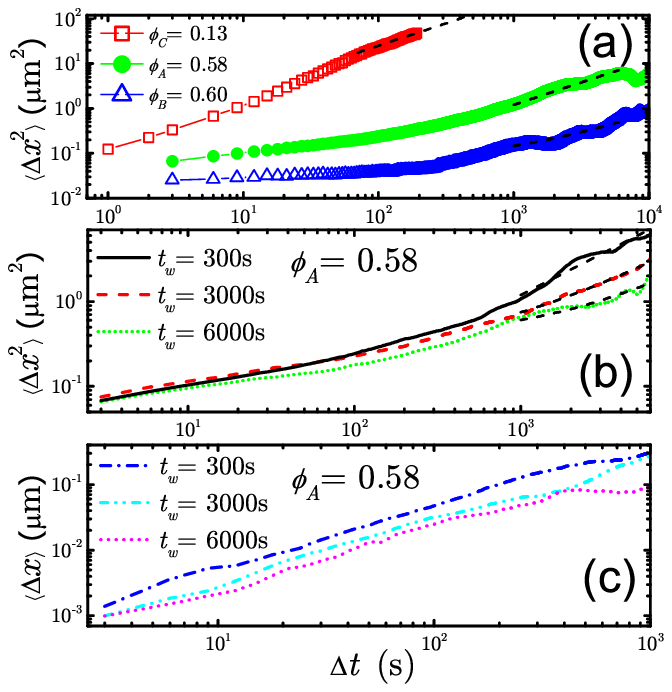}}
\caption{}
\label{msd}
\end{figure}

Figure \ref{msd}a shows $\langle\Delta x^{2}(t, t_w)\rangle$ as a
function of $\Delta t$ at a fixed $t_w =100$s calculated for the
three colloidal samples at $\phi_{C}<\phi_{g}<\phi_{A}<\phi_{B}$,
and Fig. \ref{msd}b shows the age dependence for the glassy sample
A. Sample B shows $t_w$ dependence similar to sample A, while
sample C, being at equilibrium, is independent of $t_w$, i.e. it
is stationary (see Appendix \ref{sample B and C}).


The cage dynamics is evidenced by the plateau observed in the MSD
in the two glassy samples.
The rattling of particles inside the cages is too fast ($\sim
10^{-2}s$) \cite{Megen_prl} to be observed with our visualization
capabilities. Since we focus mainly on the long relaxation time
regime, the stationary parts, $C_{st}$ and $\chi_{st}$, are
negligible compared with $C_{ag}$ and $\chi_{ag}$, in the
following we concentrate only on the aging part of the observables
and drop the subscript ${\rm ag}$: $C=C_{\rm ag}$ and
$\chi=\chi_{\rm ag}$.
The tracers' motion is confined by the cage, which persists for a
time of the order of the relaxation time $\tau(t_w)$. As expected
for an aging system, this relaxation time increases with $t_w$ as
observed in Fig. \ref{msd}b. For longer times, $\Delta t >
\tau(t_w)$, structural rearrangements lead to a second increase of
the MSD, defining a diffusion regime characterized by a diffusion
constant, $D(t_w)$, which depends on the  waiting time.
We find an
asymptotic form:

\begin{equation}\label{diffusivity}
\langle [x(t_w +\Delta t, t_w ) - x(t_w ,t_w )]^{2} \rangle \sim
2D(t_w )\Delta t, ~~~~ \mbox{for $\Delta t > \tau(t_w)$}.
\end{equation}

Figure \ref{msd}c shows the average displacement of the magnetic
beads under the external force as a function of time $\Delta t$,
for various aging times, $t_w $, in sample A. The magnetic force
is set as small as possible to observe the linear response regime,
$F=1.7\times 10^{-14}N$ (see Appendix \ref{linear}). Contrary to
the behavior of the MSD, the integrated response function does not
display the plateau characteristic of the cage effect. The data
can be fitted to:
\begin{equation}
\langle x(t_w +\Delta t, t_w ) - x(t_w ,t_w ) \rangle \sim M(t_w
)F\Delta t,
\end{equation}
where $M(t_w)$ is the mobility of the tracers which is again
waiting time dependent as seen in the figure.

\begin{figure}
\centering \resizebox{15cm}{!}{\includegraphics{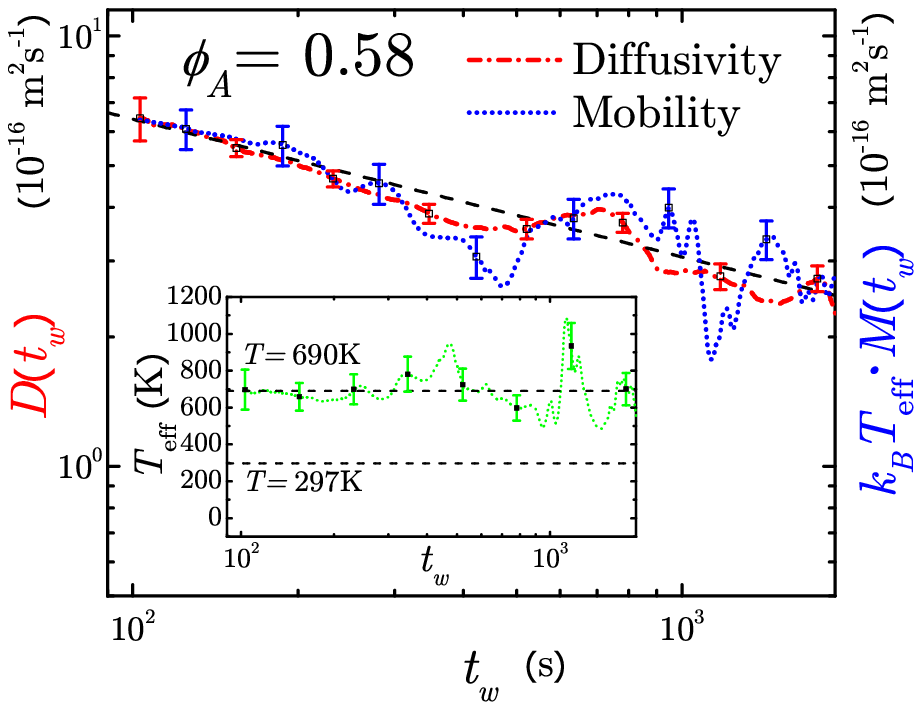}}
\caption{}
\label{d_m_scaling}
\end{figure}

To investigate the nature of the scaling behavior of the aging
regime, we study the dynamical behavior with respect to $t_w $.
Figure \ref{d_m_scaling} shows both the diffusivity and mobility
as a function of $t_w $ for sample A.
Both $D(t_w)$ and $M(t_w)$ decrease with $t_w $ signaling the
slowing down in the dynamics. More importantly, they decrease
according to a power-law with the same exponent for both
quantities:

\begin{equation}
D(t_w ) \propto {t_w }^{-\gamma} ~~~ \mbox{and} ~~~ M(t_w )
\propto {t_w }^{-\gamma}, \label{gamma}
\end{equation}
with $\gamma=0.32\pm0.08$. This result is consistent with previous
work in a similar aging colloidal system, where it was found that
the MSD $\langle \Delta x^{2} \rangle$ has a power law decay with
$t_w $ \cite{Weeks_aging}. The fact that all the quantities scale
as power laws indicates that the aging regime lasts  for a very
long time, perhaps without ever equilibrating.

{\bf Effective temperature.---} The same power law decay of the
diffusivity and mobility implies that the system has thermalized
at a  constant effective temperature $T_{\rm eff}$ independent of
$t_w $. This temperature is given by an extension of the
 Stokes-Einstein
relation or  FDT to out-of-equilibrium systems. Even though both
$D$ and $M$ depend on the age of the system, their ratio is
constant:
\begin{equation}
T_{\rm eff}(t_w ) \equiv \frac{D(t_w )}{{k_B}M(t_w )} =
(690\pm100) \mbox{K}. \label{einstein}
\end{equation}
The inset of Fig. \ref{d_m_scaling}  plots the $T_{\rm{eff}}$ as a
function of $t_w $. We obtain $T_{\rm eff}\approx 690$K which is
more than double the ambient temperature of 297K. For very large
 $t_w $, $T_{\rm{eff}}$ shows large fluctuations that
are mainly due to the larger statistical error (due to the limited
number of tracers) in obtaining the mobilities and diffusivities
in the large $\Delta t$ and $t_w$ regime. It remains  a question
whether the long waiting time regime may show interrupted aging.


Equation (\ref{einstein}) is easy to understand when the system is
at equilibrium: we extract energy from many identical tracers
located in distant regions of the colloidal system
and transfer it to the thermometer system. The thermometer
receives work from the diffusive motion of the tracers, and
it  dissipates energy through the viscosity of the system. These
two opposing effects  make the  thermometer stabilize at a
temperature guaranteed by the Einstein relation. Naturally, we
have applied the diffusion-mobility calculations to dilute sample
C and find that it is equilibrated at the bath temperature (see
Appendix \ref{sample B and C}).
On the other hand, the colloidal sample is aging out of
equilibrium. Nevertheless, the fact that the ratio of diffusion to
mobility yields a constant temperature can be  taken
 as an indication that the long time behavior of the
system has thermalized at a larger effective temperature $T_{\rm
eff}\approx 690$K.

Although it may seem counterintuitive that the slow relaxation at
long times corresponds to a temperature that is actually higher
than the equilibrium bath temperature, there is an interesting
physical picture that rationalizes this observation. One can think
of the energy landscape of configurations of the colloidal glass
being explored less frequently, yet the amplitude of the jumps
between basins corresponds to ``hotter" explorations of a boarder
distribution of energy states. While this mechanism violates the
usual relations between particle motion and temperature, it gives
rise to the effective temperature measured in our experiments.



\begin{figure}
\centering \resizebox{15cm}{!}{\includegraphics{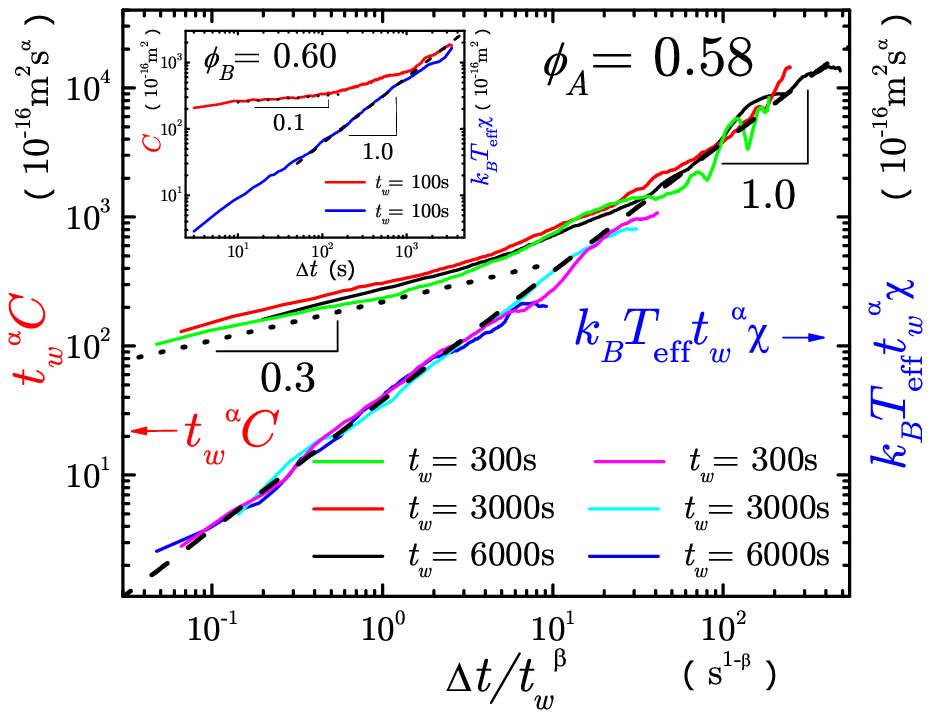}}
\caption{} \label{scaling}
\end{figure}

{\bf Scaling ansatz for the global correlations and responses.---}
Further insight into the understanding of the slow relaxation can
be obtained from the study of the universal dynamic scaling of the
observables with $t_w$. Based on spin-glass models, different
scaling scenarios have been proposed \cite{Cugliandolo, Henkel}
for correlation and response functions. Our analysis indicates
that the observables can be described as

\begin{subequations}\label{scaling-laws}
\begin{align}
C(t_w +\Delta t, t_w )&=\langle \Delta x^{2} \rangle/2 =
t_w^{-\alpha}f_{D}(\frac{\Delta
t}{{t_w }^{\beta}}),\label{scaling-laws1}\\
\chi(t_w +\Delta t, t_w )&=\frac{\langle \Delta x \rangle}{F} =
t_w^{-\alpha}f_{M}(\frac{\Delta t}{{t_w
}^{\beta}}),\label{scaling-laws2}
\end{align}
\end{subequations}
where $f_{D}$ and $f_{M}$ are two universal functions and $\alpha$
and $\beta$ are the aging exponents. Evidence for the validity of
these scaling laws is provided in Fig. \ref{scaling} where the
data of the correlation function and the integrated response
function collapse onto a master curve when plotted as $
t_w^{\alpha}C(t_w +\Delta t,t_w )$ and $t_w^{\alpha}\chi(t_w
+\Delta t,t_w )$ versus $\Delta t/{t_w }^{\beta}$. By minimizing
the $\sigma^2$ value of the difference between the master curve
and the data (see Appendix \ref{exponents}) we find that the best
data collapse is obtained for the following aging exponents:
$\alpha + \beta = 0.34\pm0.05$ and $\beta = 0.48\pm0.05$. We find
(Fig. \ref{scaling}) that the scaling functions satisfy the
following asymptotic behavior:
\begin{subequations}
\begin{align}
f_{D}(y)&\sim \{\begin{array}{lll}y^{0.3}&\ y \ll 1,\\y&\ y \gg 1,
\end{array}
\\
f_{M}(y)&\sim y,
\end{align}
\end{subequations}
in agreement with the fact that the motion of the particles is
diffusive at long times,
 $\langle \Delta x^{2} \rangle \sim \Delta t$,
and the existence of a well-defined mobility, respectively.
Therefore at long times, both the correlation and response
functions display the same power law decay:
\begin{subequations}
\begin{align}
C(t_w +\Delta t,t_w ) & \sim  t_w^{-(\alpha+\beta)} \Delta t,\\
\chi(t_w +\Delta t,t_w ) & \sim  t_w^{-(\alpha+\beta)} \Delta t.
\end{align}
\end{subequations}

The result $C(t_w +\Delta t, t_w) \sim \chi(t_w+\Delta t, t_w)
\sim t_w^{-0.34}$ confirms our previous result, Eq. (\ref{gamma}),
$D(t_w )\sim M(t_w) \sim t_w ^{-0.32}$. This is in further
agreement with the finding that $T_{\rm eff}$ is independent of
the age of the system, $t_w$. For short times, the MSD scaling
function crosses over to a sub-diffusive behavior of the
particles. We obtain,  $\langle x^2(t_w +\Delta t,t_w )\rangle
\sim t_w^{-\alpha}(\Delta t/t_w^{\beta})^{0.3}=t_w^{-0.004}\Delta
t^{0.3}$. Since the trapping time corresponds to the size of the
cages denoted by $q(t_w)$, we can determine the cage dependence on
$t_w$ as $q(t_w)\sim t_w^{-0.002}$. The resulting exponent is so
small that we can say that the cages are not evolving with the
waiting time, within experimental uncertainty. Furthermore, the
scaling ansatz of Eq. (\ref{scaling-laws}) indicates that the
relaxation time of the cages scales as $\tau(t_w)\sim t_w^{\beta}$
since this is the time when the sub-diffusive behavior
crosses-over to the long time diffusive regime.

From Fig. \ref{scaling} we see that the asymptotic $\Delta
t$-linear regime appears when the reduced variable $\Delta
t/{t_w}^{\beta}$ is larger than 10, $\Delta t/{t_w}^{\beta} > 10$.
The time separation between $\Delta t$ and $t_w$ can be determined
using the cut-off: $\Delta t/{t_w}^{0.48} \sim 10$. For smaller
times, $\Delta t/{t_w}^{\beta} < 10$, we obtain a subdiffusion
regime (with exponent 0.3), which is characteristic of the cage
dynamics. For longer times, $\Delta t/{t_w}^{\beta} > 10$, we
observe the asymptotic $\Delta t$-linear regime where the
diffusivity is calculated. The measurement of MSD in Fig. 1
extends up to $t_w = 6000$s. For this $t_w$, the separation of
time scales appears when $\Delta t > 10*6000^{0.48} \simeq 10^3$s.
Our measurements for MSD extend to $10^4$s, ensuring a separation
of time scales even for this longest waiting time. For the smaller
$t_w$ considered in the calculation of $T_\texttt{eff}$, the
separation of time scales is even more pronounced. For instance,
for a typical $t_w=1000$s where the $T_{\texttt{eff}}$ is
calculated, the separation of time scales occurs at
$10*1000^{0.48} \simeq 275$s, again ensuring a well defined long
time asymptotic behavior for $\Delta t = 10^4$s.

Although the existence of an effective temperature can be
rationalized using theoretical frameworks of disordered spin-glass
models \cite{Kurchan_prl}, we find that the scaling forms of the
correlations and responses are not consistent with such models.
Based on invariance properties under time reparametrization,
spin-glass models predict a general scaling form $C_{\rm ag}(t,t_w
) = C_{\rm ag}(h(t)/h(t_w))$, where $h(t)$ is a generic monotonic
function \cite{Cugliandolo}. We find that the scaling of our
observables from Eqs. (\ref{scaling-laws1}), (\ref{scaling-laws2})
cannot be collapsed with the ratio $h(t)/h(t_w)$. The scaling with
$h(t)/h(t_w)$ is expected for system in which the correlation
function saturates at long times \cite{Cug_Dou}. On the other
hand, our system is diffusive, and the studied correlation
function is not bounded. Indeed, similar scaling as in our system
has been found in the aging dynamics of anther unbounded system:
an elastic manifold in a disordered media \cite{Bustingorry}. The
suggestion is that this problem and the particle diffusing in a
colloidal glass may belong to the same universality class.
Furthermore our results can be interpreted in terms of the droplet
picture of the aging of spin glass, where the growth of the
dynamical heterogeneities control the aging.

{\bf Local fluctuations of autocorrelations and responses.---}
Previous work has revealed the existence of dynamical
heterogeneities, associated with the cooperative motion of the
particles, as a precursor to the glass transition as well as in
the glassy state \cite{Kegel,Weeks_sci,Weeks_aging,Kob}. Instead
of the average global quantities studied above, the existence of
dynamical heterogeneities requires a microscopic insight into the
structure of the glassy. Earlier studies focused mainly on
probability distributions of the particles displacement near the
glass transition. More recent analytical work in spin glasses
\cite{Castillo} shows that the probability distribution function
(PDF) of the local correlation $P(C)$ and the local integrated
response $P(\chi)$ could reveal essential features of the
dynamical heterogeneities.

\begin{figure}
\centering \resizebox{9cm}{!}{\includegraphics{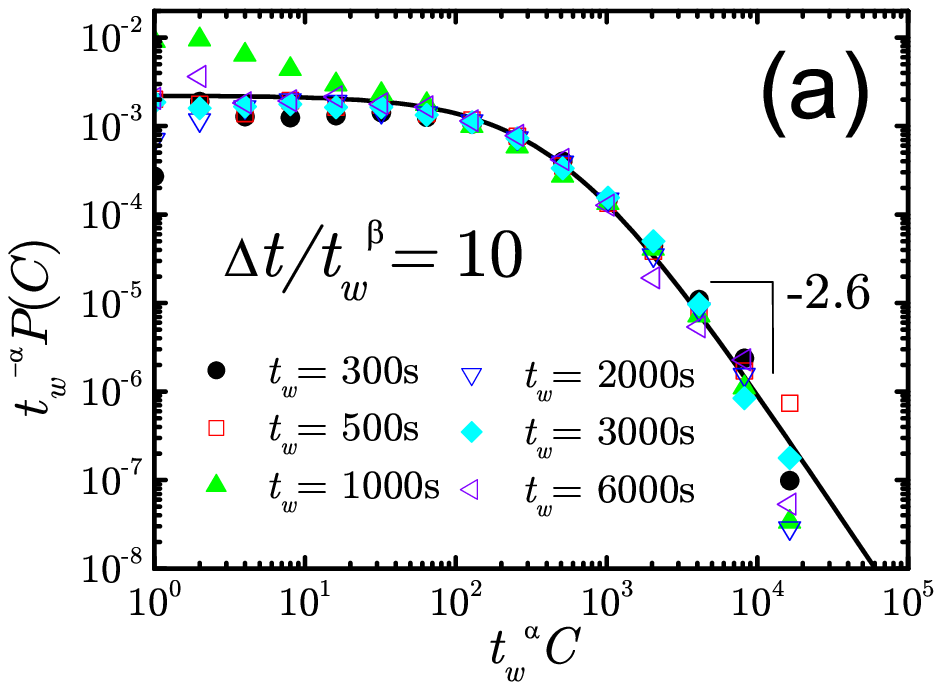}}%
\centering \resizebox{9cm}{!}{\includegraphics{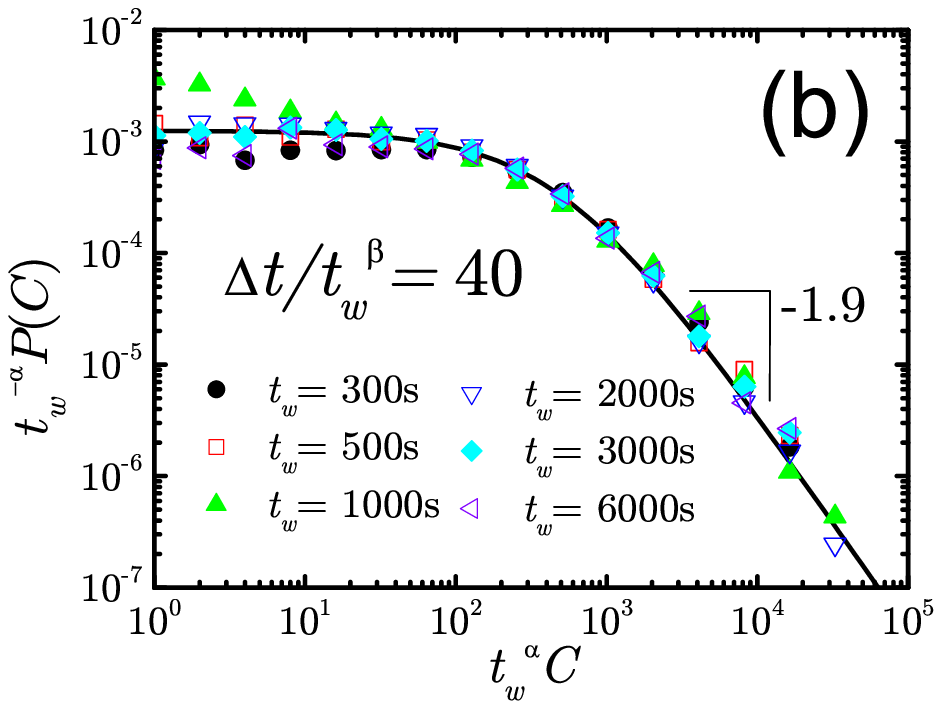}}
\centering \resizebox{9cm}{!}{\includegraphics{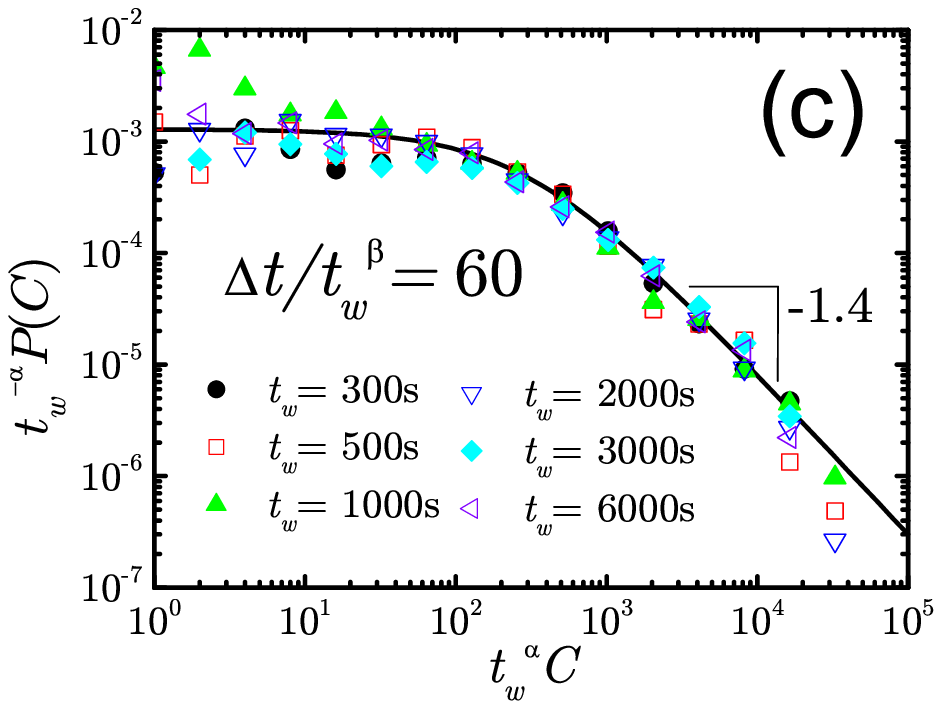}}%
\centering \resizebox{9cm}{!}{\includegraphics{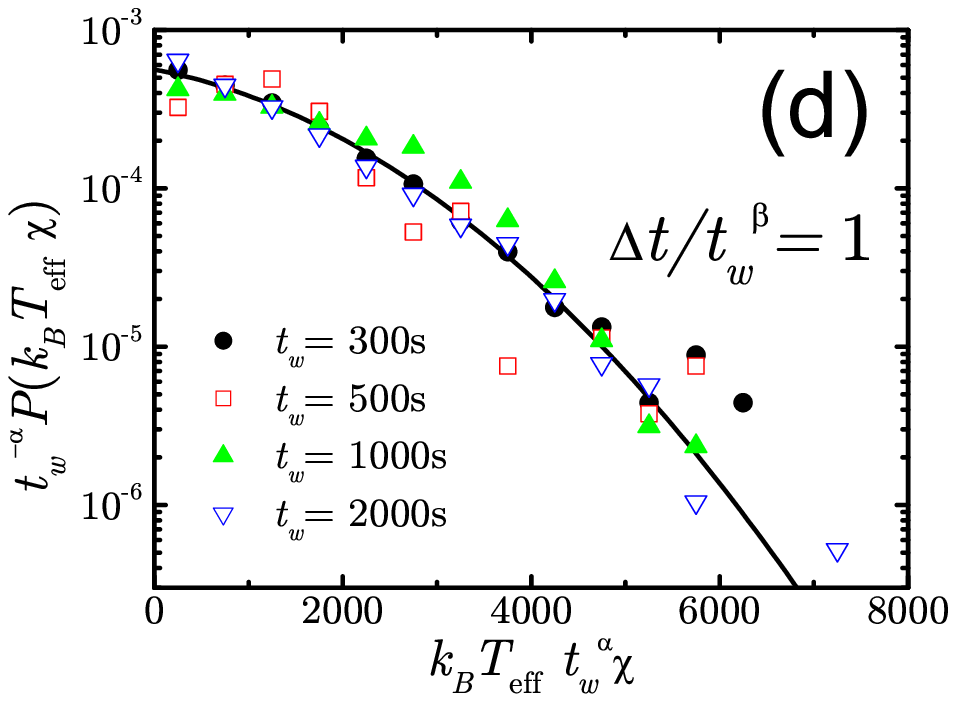}}
\caption{} \label{pdf}
\end{figure}

Here we perform a systematic study of $P(C)$ and $P(\chi)$ in
sample A, and the resulting PDFs are shown in Fig. \ref{pdf}.
The scaling ansatz of Eq. (\ref{scaling-laws}) implies that $P(C)$
and $P(\chi)$ should be collapsed by rescaling the time $\Delta t$
by $t_w^{\beta}$ and the local fluctuations by $t_w^{\alpha}$ (see
Appendix \ref{local} for more details). Indeed, this scaling
ansatz provides the correct collapse of all the local fluctuations
captured by the PDFs, as shown in Fig. \ref{pdf}a, \ref{pdf}b and
\ref{pdf}c for $P(C)$ and in Fig. \ref{pdf}d for $P(\chi)$.

The PDF of the autocorrelation function displays a universal
behavior following a modified power-law
$t_w^{-\alpha}P(C)\propto(t_w^\alpha C+C_0)^{-\lambda}$, where
$C_0$ and $\lambda$ only depend on the time ratio $\Delta
t/t_w^\beta$. For the smaller values of $C$ ($C<t_w^{-\alpha}
C_0$), the existence of a flat plateau in $P(C)$
indicates that the tracers are
confined in the cage. For larger values of $C$, the salient
feature of the PDF is the very broad character of the
distribution, with an asymptotic behavior $P(C)\sim
C^{-\lambda}$. This large deviation from a Gaussian
behavior is a clear indication of the heterogeneous character of
the dynamics. Furthermore, the exponent $\lambda$ decreases from
2.6 to 1.4 with the time ratio $\Delta t/t_w^{\beta}$ ranging from
10 to 60. We notice that $\lambda =2$ corresponds to the
crossover between the short-time and long-time regime in Fig.
\ref{scaling}, where $\Delta t/t_w^\beta\approx40$. The
significance of $\lambda=2$ is seen in the integral $\int
P(C)CdC$. For $\lambda>2$
($\Delta t/t_w^\beta < 40$) the plateau dominates over the
power law tail in the integral and the dynamics is less heterogeneous. For
$\lambda<2$ ($\Delta t/t_w^\beta > 40$)
the power law tail dominates and this regime
corresponds to the highly heterogeneous long-time regime (see
Appendix \ref{power law}).

On the contrary, $P(\chi)$, shown in Fig. \ref{pdf}d, displays a
different behavior. The fluctuations are more narrow and the PDF
can be approximated by a Gaussian. This is consistent with the
fact that we did not find cage dynamics for the global response in
Fig. \ref{msd}c. Moreover, numerical simulations of spin-glass
models \cite{Castillo} seem to indicate a narrower distribution as
found here.


We have presented experimental results on an aging colloidal glass
showing a well-defined temperature for the slow modes of
relaxation of the system. This $T_{\rm eff}$ is larger than the
bath temperature since it implies large scale structural
rearrangements of the particles. In other words, it controls the
cooperative motion of particles needed to relax the cages. The
interesting result is that this temperature remains constant
independent of the age, even though both the diffusivity and the
mobility are age dependent. The power-law scaling found to
describe the transport coefficients indicates the slow relaxation
of the system. A universal scaling form is found to describe all
the observables. That is, not only the global averages, but also
the local fluctuations. The scaling ansatz, however, cannot be
described under present models of spin-glasses, but it is more
akin to that observed in elastic manifolds in random environments
suggesting that our system may share the same universality class.

\newpage

\centerline{\bf Method: Experimental details} The colloidal
suspension is immersed in a solution  containing $76\%$ weight
fraction of cyclohexylbromide and $24\%$ cis-decalin which are
chosen for their density and index of refraction matching
capabilities \cite{Weeks_sci}. For such a system the glass
transition occurs at
$\phi_{g}\approx 0.57 - 0.58$ \cite{Pusey,Kegel,Weeks_sci}.
In our experiments we consider three samples at different
densities and determine the glassy phase for the samples that
display aging. The main results are obtained for sample A just
above the glass transition $\phi_{\rm A}= 0.58\pm0.01$. We also
consider a denser sample B with $\phi_{\rm B}=0.60\pm0.01$,
although this sample is so deep in the glassy phase that we are
not able to study the slow relaxation of the system and the
dependence of the waiting time within the time scales of our
experiments.
Finally, we also consider a sample C below the glass transition
$\phi_{\rm C}=0.13\pm0.01 < \phi_g$ for which we find the usual
equilibrium relations (see below). Prior to our measurements, the
samples are homogenized by stirring them for two hours to achieve
a reproducible initial time (see Appendix \ref{experiment} for a
full discussion).

We use a magnetic force as the external perturbation to generate
two-dimensional motion of the tracers \cite{Song} (see Appendix
\ref{experiment}) on a microscope stage following a simplified
design of \cite{magnetic_stage}. Video microscopy and computerized
image analysis are used to locate the tracers in each image. We
calculate the response and correlation functions in the x-y plane.


\begin{acknowledgments}
We wish to thank M. Shattuck for help in the design of the
experiments and J. Bruji\'c and S. Mistry for providing the colloidal
particles and illuminating discussions. We acknowledge the
financial support from DOE and NSF.
Correspondance should be addressed to H. A. Makse.

\end{acknowledgments}

\clearpage

FIG. \ref{msd}. Autocorrelation and response functions. (a) Mean
square displacement of tracers as a function of $\Delta t$ for the
three samples $\phi_{C}=0.13<\phi_{g}<\phi_{A}=0.58<\phi_{B}=0.60$
at $t_w =100$s. The straight dash lines indicate estimates for the
diffusivity $D(t_w)$. (b) Mean square displacement of tracers in
sample A as a function of time $\Delta t$, for various aging times
$t_w $. The dashed straight lines indicate the fitting regime to
calculate the diffusivity $D(t_w)$. (c) Mean displacement of
tracers under the magnetic force in sample A for various aging
times $t_w $.

FIG. \ref{d_m_scaling}. Diffusivity (red dash-dot line) and
mobility (blue dot line) as a function of aging time for sample A.
A straight dash line is added to guide the eyes, which shows
$D(t_w ) \sim {t_w }^{-\gamma}$ and $M(t_w ) \sim {t_w
}^{-\gamma}$ with $\gamma=0.32\pm 0.08$. For convenience of
comparison with diffusivity, mobility is scaled by $k_BT_{\rm
eff}$, with $k_B$ the Boltzmann constant and the effective
temperature $T_{\rm eff}=690K$. The inset shows the effective
temperature as a function of waiting time. Error bars are added
only to some representative points for clarity.

FIG. \ref{scaling}. Scaling plot of sample A for the scaled
autocorrelation, ${t_w }^{\alpha}C$, and scaled integrated
response, $k_{B}T_{\rm eff}{t_w }^{\alpha}\chi$, as a function of
the time ratio, $\Delta{t}/{t_w }^{\beta}$, for different waiting
times. The black dash line is a linear fit which indicates that
$T_{\rm eff}=690K$. The inset is a plot of sample B for
autocorrelation, $C$, and integrated response, $k_{B}T_{\rm
eff}\chi$, as a function of $\Delta{t}$ at $t_{w}=100s$. The black
dash line is a linear fit which indicates that $T_{\rm
eff}=1600K$.

FIG. \ref{pdf}. (a) PDF of the scaled local correlation
$t_w^{\alpha}C$ for $\Delta t/t_w ^{\beta}=10$ with $\Delta t$
from 155s to 650s. Solid line corresponds to the modified
power-law fit $t_w^{-\alpha}P(C)\propto(t_w^\alpha
C+C_0)^{-\lambda}$, with $C_0=500$ and $\lambda=2.6$. (b) Same as
(a), for $\Delta t/t_w ^{\beta}=40$ with $\Delta t$ from 620s to
2600s. The resulting distribution can be fitted (solid line) by a
modified power law with $C_0=500$ and $\lambda=1.9$. (c) Same as
(a), for $\Delta t/t_w ^{\beta}=60$ with $\Delta t$ from 930s to
3900s. The solid line is the best fit, with parameters $C_0=300$
and $\lambda=1.4$. (d) PDF of the scaled local integrated response
for $\Delta t/t_w ^{\beta}=1$ shows a Gaussian decay. For the
convenience of the comparison with (a), (b), (c), we rescale the
response by $k_BT_{\rm eff}$, where $T_{\rm eff} = 690$K.

\clearpage

\clearpage

\clearpage

\clearpage

\clearpage

\appendix{ \bf SUPPLEMENTARY MATERIALS}

\section{Details of the experimental set up}
\label{experiment}

\begin{figure}
\centering \resizebox{12cm}{!}{\includegraphics{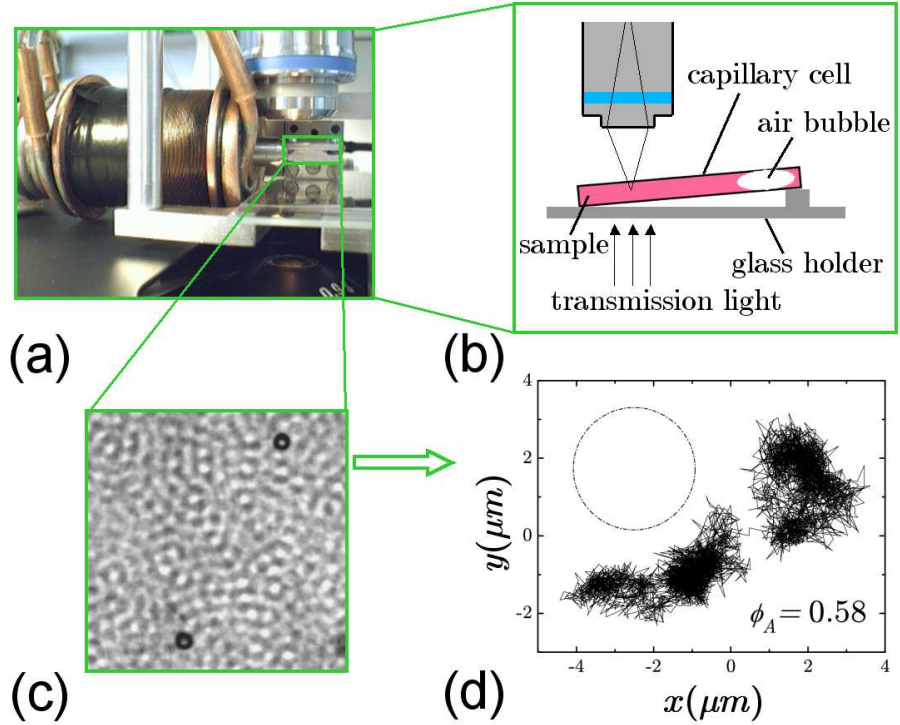}}
\caption{(a) Picture of the experimental setup. (b) Schematic
illustration of the setup. (c) Detail of an image of tracer in
sample A. (d) Trajectory of a tracer in sample A showing the cage
dynamics over 4 hours. The circle represents the size of the
tracer particle, $3.2\mu m$.} \label{images}
\end{figure}

The experimental set up is shown in Fig. \ref{images}a and Fig.
\ref{images}b. We use a Zeiss microscope with a 50$\times$
objective of numerical aperture 0.5 and a working distance of 5mm.
We work with a field of view $194\mu m\times155\mu m$ by using a
digital camera with the resolution of 1288 pixels $\times$ 1032
pixels. We locate the center of each bead position with sub-pixel
accuracy, by using image analysis. For the condensed samples A and
B, we use the low frame rate $1/3$ frame/sec, while for the dilute
sample C, we record the images at $1$ frame/sec. The long working
distance of the objective is necessary to allow the pole of the
magnet to reach a position near the sample. An example of the
images of the tracers obtained in sample A is shown in Fig.
\ref{images}a where we can see the black magnetic tracers embedded
in the background of nearly transparent PMMA particles. An example
of the trajectory in the x-y plane of a tracer diffusing without
magnetic field is shown in Fig. \ref{images}d. We note that this
particular tracer moves away from two cages in a time of the order
of $4$ hours.

The magnetic field is produced by one coil made of 1200 turns of
copper wire. We arrange the pole of the coil perpendicular to the
vertical optical axis, and generate a field with no vertical
component. Thus, the tracers move in the x-y plane with a slight
vertical motion which is generated by a density mismatch between
the tracers and the background PMMA particles. This vertical
motion is very small at the high volume fraction of interest here,
and therefore we calculate all the observables in the x-y plane.

The magnetic force is calibrated for a given coil current by
replacing the suspension with a mixture of $50:50$ water-glycerol
solution with a few magnetic tracers. The distance between the top
of the magnetic pole and the vertical optical axis is always
fixed, which means that the magnetic force at the local plane
depends only on the coil current. At a given current, we determine
the velocity of the magnetic tracers at the focal point and
calculate the magnetic force from Stokes's law, $F = 6\pi\eta
a_{m}v$, where $\eta$ is the viscosity of the water-glycerol
solution, $a_{m}$ is the tracer radius, and $v$ is the observed
velocity of the tracer. The uncertainty in the obtained force
comes from: (a) the uncertainty in the coil current which is
$1\%$, (b) the beads, which are not completely monodisperse in
their magnetic properties, causes a $10\%$ uncertainty
\cite{Weeks_force}, and (c) the magnetic field is slowly decaying
in the field of view, causing a $4\%$ uncertainty.


\section{Sample Preparation at $t_w=0$}
\label{inistate}

\begin{figure}
\centering \resizebox{12cm}{!}{\includegraphics{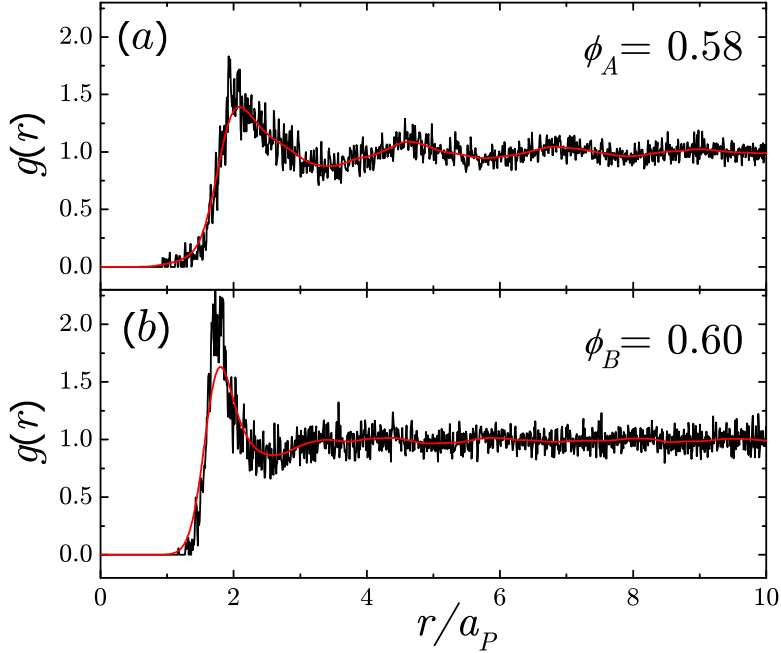}}
\caption{Pair-distribution functions of (a) sample A and (b)
sample B.}\label{gr}
\end{figure}

It is important to correctly determine the initial time of each
measurement to reproduce the subsequent particle dynamics. We
initialize the system by stirring the sample for two hours with an
air bubble inside the sample (see Fig. \ref{images}b) to
homogenize the whole system and break up any pre-existing
crystalline regions. Then we place the sample on the magnetic
stage and take images to obtain the trajectories of the tracers
that appear in the field of view. The initial times $t = t_w = 0$
are defined at the end of the stirring. After measuring all the
tracers appearing in the field of view, a new stirring is applied,
the waiting time is reset to zero, and the measurements are
repeated for a new set of tracers. We have analyzed the
pair-distribution function, $g(r)$, and two-time intensity
autocorrelation function, $g_{2}(t_w,t_w+ \Delta t)$, in order to
test our rejuvenation technique.

First, Fig. \ref{gr} plots the pair-distribution functions of
sample A and sample B right after the stirring procedure. We
calculate the pair-distribution functions by reconstructing the
packings from 3D confocal microscopy images of size $60\mu
m{\times}60\mu m{\times}15\mu m$. We find that the samples do not
show obvious crystallized region by directly looking either at the
pair-distribution function or the images taken from confocal
microscopy.
 For these measurements we use a Leica
confocal microscope.
The PMMA particles are fluorescently dyed so that they are ready
to be observed by confocal microscopy.
We load the samples
sealed in a glass cell on the confocal microscope stage
 and use a  Leica
HCX PL APO  63x, 1.40 numerical aperture, oil immersion lens
for 3D particle visualization in order
 to calculate the volume fraction and the pair-distribution
function.

\begin{figure}
\centering \resizebox{12cm}{!}{\includegraphics{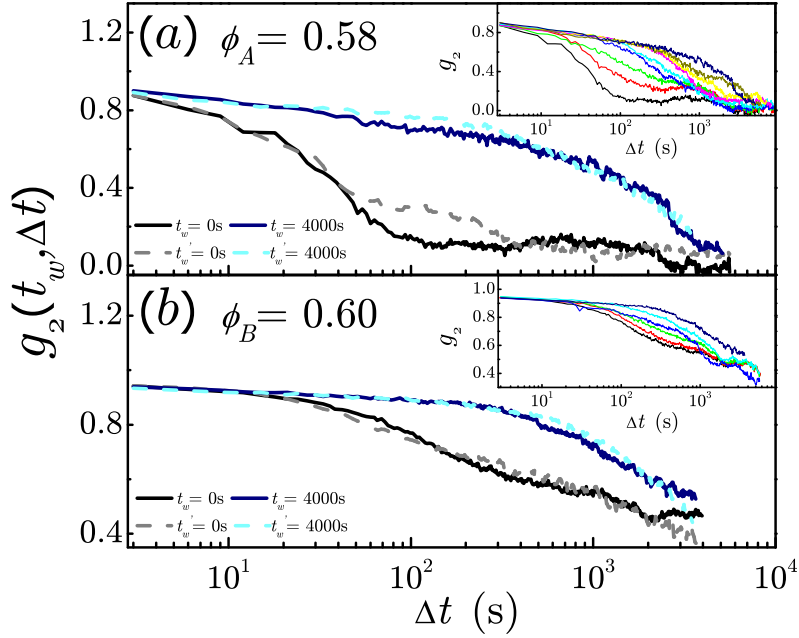}}
\caption{Two-time intensity autocorrelation function, $g_2(t_w+
\Delta t, t_w)$, of (a) sample A and (b) sample B. The solid black
and navy curves correspond to $t_w=0$s and $t_w=4000$s,
respectively. The dash gray and light blue curves correspond to
the measurements after the second stirring with ${t_w}^{'}=0$s and
${t_w}^{'}=4000$s, respectively. The fact that the curves at
$t_w=0$s and ${t_w}^{'}=0$s coincide indicates that sample has
been fully rejuvenated to its initial state. The inset is the same
plot as the main figure for different $t_w$ ranging from 0s (solid
black) to 4000s (solid navy). }\label{g2}
\end{figure}

Second, we analyze $g_2(t_w+ \Delta t, t_w)$ and show that the
sample has been rejuvenated by the stirring process. We plot
$g_2(t_w+ \Delta t, t_w)$ for 3 situations: (a) after the first
stirring process ($t_w=0$), (b) after the sample has aged for a
long time ($t_w=4000s$), (c) we then apply the stirring again and
immediately plot the autocorrelation function for the new initial
time (${t_w}^{'}=0$). The plot show the lack of correlation for
the two individual measurements $t_w=0$ and ${t_w}^{'}=0$, thus
demonstrating that the sample has been rejuvenated. Technically,
we record the temporal image sequence of sample A and sample B,
and study the aging by calculating two images' correlation defined
as:

\begin{equation}
g_{2}(t_w,t_w+ \Delta t)=\frac{\langle I(t_w)I(t_w+\Delta
t)\rangle}{\langle I(t_w)^2\rangle},
\end{equation}
where $I$ is the average gray-scale intensity of a small box (20
pixels $\times$ 20 pixels, PMMA particles' diameter is roughly
equal to 20 pixels), and $I(t_w)$ and $I(t)$ come from the images
at different times $t_w$ and $t$, respectively. We cut one image
(1288 pixels $\times$ 1032 pixels) into many boxes, and calculate
the correlation of two boxes located at the same position in the
two images. The average $\langle...\rangle$ is taken over all the
boxes in one image. Fig. \ref{g2} plots the correlation function,
$g_{2}(t_w,t_w+ \Delta t)$, of sample A and sample B, which shows
aging behavior (see the inset). Fig. \ref{g2} shows how
$g_{2}(t_w,t_w+ \Delta t)$ calculated after two stirring processes
at $t_w=0$ and ${t_w}^{'}=0$ coincide, indicating that the sample
can be fully rejuvenated to its initial state by our stirring
technique.

\section{Data analysis}
\label{data}

The average $\langle\cdot\cdot\cdot\rangle$  used to calculate the
MSD $\langle\Delta x^{2}(t ,t_w)\rangle$ and the mean displacement
$\langle\Delta x(t ,t_w)\rangle$ denotes an ensemble average over
the tracers and over the initial time $t_{0}$ which is varied over
a small time interval centered at $t_w $. We choose a small
coarse-grained time interval compared to $t_w $, so that we can
ignore the aging effect in  such short time. We then regard all
the trajectories in this interval region as having the same
waiting time $t_w $. In practice, at a given $t_w $ and a given
$\Delta t=t-t_w =t_{1}-t_{0}$, $(t>t_w ,t_{1}>t_{0})$, we use the
common method
(\cite{window_average}, pages 118-122) of opening a
small time window $[t_w -t_{r},t_w +t_{r}]$ and perform an average
over all the $\Delta t$ with $t_{0}\in[t_w -t_{r},t_w +t_{r}]$ and
$t_{1}\in[t_w -t_{r}, t_{max}]$ in order  to measure the transport
coefficients. We note that  $t_{r} < t_w $ and that $t_{max}$ is
the maximum time of our measurements. The diffusion and mobility
constants are then obtained by averaging not only over the tracers
but also over the initial time $t_{0}$. This common technique
allows us to obtain an estimation of the diffusion constant and
mobility constant by using 82 tracers and 75 tracers respectively
for this particular system. We check that by reducing the region
size $t_{r}$ we obtain the same behavior of $D(t_w )$ and $M(t_w
)$, so that the transport coefficients are independent of the
averaging technique.

We collect all the tracers' trajectories from several measurements
since our system is well reproducible as described in Appendix
\ref{experiment}. In each measurement, we consider tracers that
are separated by at least 10 particle diameters from the sample
boundary and from each other. This guarantees that the
tracer-tracer magnetic interaction can be ignored.

\section{$\Delta t$-linear regime}

\begin{figure}
\centering \resizebox{12cm}{!}{\includegraphics{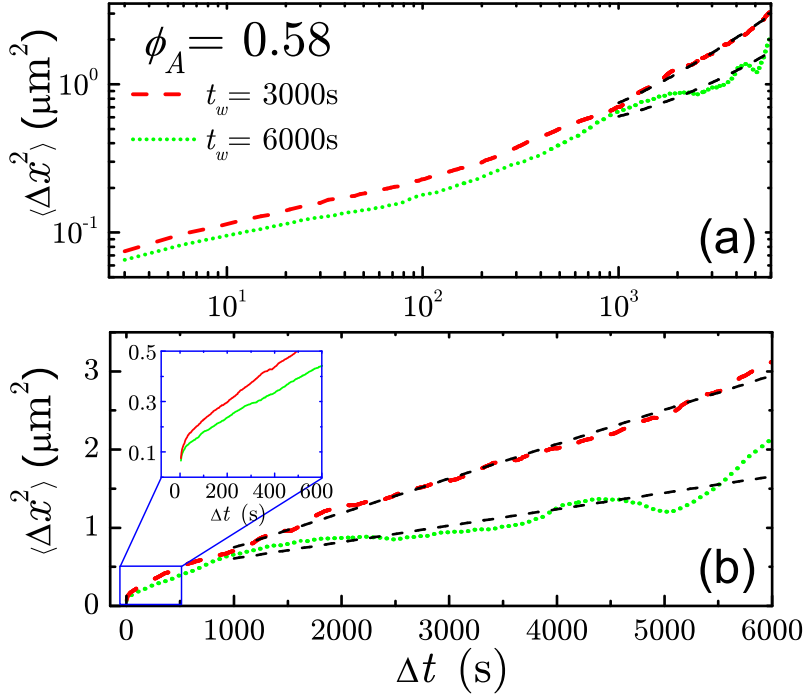}}
\caption{ Mean square displacement of tracers as a function of
$\Delta t$ for the sample A at $t_w = 3000$s and $t_w = 6000$s in
(a) log-log plot and (b) linear-linear plot. The dashed lines are
linear fittings of the $\Delta t$-linear regime.}
\label{linear_regime}
\end{figure}

In this appendix we address the issue of the proper
determination of the long $\Delta t$-linear regime in the MSD where
the diffusivity is calculated. The crossover from the cage
dynamics regime to the long time diffusive regime seems to be very
gradual in Fig. \ref{msd}b (see for instance the case of
$t_w=6000$s). We notice that due to the existence of the plateau
in the MSD, the fitting to the long
$\Delta t$-linear regime is of the form $\langle \Delta x^2\rangle
\sim A(t_w) + 2 D(t_w) \Delta t$, where $A(t_w)$ is a constant
dependent only on $t_w$ and $D(t_w)$ is the diffusion constant.
Thus, in the log-log plot of Fig. \ref{msd}, such a linear
relation would not be a complete straight line. [note that the
logarithmic plot is necessary to observe the different time scales
in the experiment, and the fitting constant $A(t_w)$ is  the
y-intercept of the fitting of the long $\Delta t$-linear regime,
for which there is no particular physical meaning].

In order to address the issue of the existence of the long-time
linear regime we compare a log-log plot of the MSD with a
linear-linear plot of the same quantity for the sample A in Fig.
\ref{linear_regime}a and Fig. \ref{linear_regime}b, respectively.
We check our results for $t_w = 3000$s and $t_w = 6000$s.

First we note that the log-log plot of Fig. \ref{linear_regime}a
naturally emphasizes the short time scales of the cage dynamics.
Thus, a linear fitting function plus a constant,
 $\langle\Delta x^2\rangle=A(t_w)+ 2 D(t_w) \Delta t$ would be represented as a smoothly
varying curve for long $\Delta t$ in such a log-log plot, giving
the impression that the long-time linear regime is not
well-defined. In order to  determine more clearly the large time
regime we have plotted the MSD in a linear-linear plot in Fig.
\ref{linear_regime}b. The $\Delta t$-linear regime and the fitting
region can be more clearly seen from the linear-linear plot. The
smaller $t_w$ are also plotted in the figure ($t_w=3000$s). As
discussed above the linear-linear plot shows more clearly the
existence of the linear regime while the log-log plot shows the
existence of the cage dynamics regime.

We also note that the fast motion inside the cage is determined by
the viscosity of the surrounding fluid, and not by the macroscopic
viscosity which determines the $T_\texttt{eff}$ at long times.
Indeed, the motion inside the cage is supposed to be equilibrated
at room temperature and satisfies the regular FDT according to
previous studies, for instance see simulations of Lennard-Jones
liquids \cite{Barrat_Kob}. The time scales for this microscopic
viscosity to dominate is of the order of $10^{-2}$s, as obtained
in light scattering experiments of colloidal glasses
\cite{Megen_prl}. These time scales are too fast to be studied
with our visualization schemes. Our measurements start at $\Delta
t \sim 1$s and go on up to $10^4$s. In this long time regime the
microscopic viscosity does not dominate the dynamics and indeed we
find the $T_\texttt{eff}$ is dominated by the macroscopic
viscosity given by the particles.

From Stokes-Einstein relation, $F=6\pi a_m \eta \upsilon$, with $F
\sim 10^{-14}N$, $\eta \sim 10^{-3}N.s.{m}^{-2}$ the viscosity of
the decalin-bromide solution, $a_m \sim 3\mu m$ the tracer's
diameter and $\Delta t \sim 10^{-2}$s, we find $\Delta x \sim
0.002\mu m$. At this length scale the particle experience the
viscosity of the surrounding liquid alone. Since we measure much
larger displacements, we conclude that the tracers are
experiencing the macroscopic viscosity and the effective
temperature is therefore well defined.

Therefore, we believe that we have achieved the necessary
separation of length scales which allows us to define a
$T_\texttt{eff}$. The small displacements observed in the response
function are due to the fact that the magnetic forces ought to be
small enough to achieve the linear response regime. The small
displacements obtained in our measurements are consistent with
previous confocal microscopy studies of colloidal suspensions near
the glass transition \cite{Weeks_prl}, which found motions of the
order of $1/10$ of the particle diameter.

\section{Study of samples C}
\label{sample B and C}

\begin{figure}
\centering \resizebox{12cm}{!}{\includegraphics{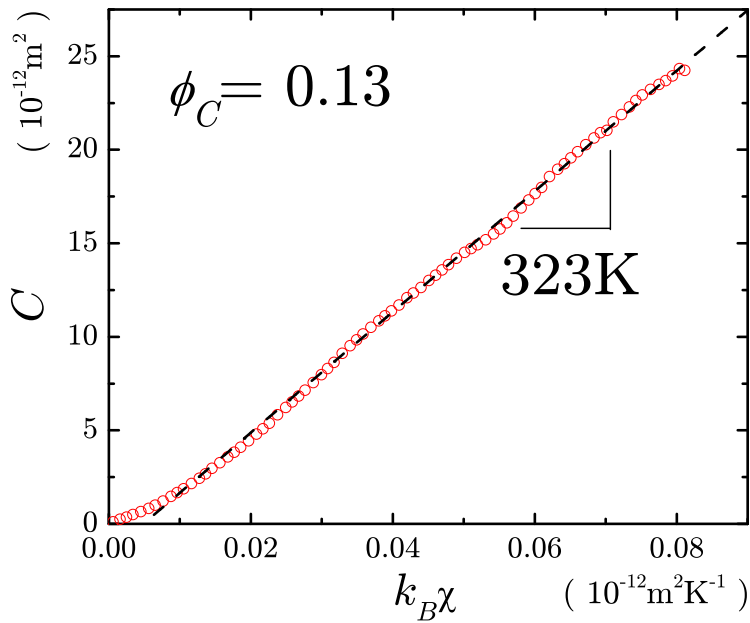}}
\caption{ Parametric plot of correlation versus integrated
response function for dilute sample C. The fitting line shows that
the temperature of sample C, $T=(323\pm30)$K, is close to the bath
temperature ($T_{\rm bath}=297$K) in our experiment.}
\label{sample C}
\end{figure}

Figure \ref{sample C} shows the parametric plot of correlation
function $C$ versus integrated response function $k_{B}\chi$ for
the dilute sample C, $\phi_C=0.13$. We find
$M=3.28\times10^7$ms$^{-1}$N$^{-1}$ and
$D=1.46\times10^{-13}$m$^{2}$s$^{-1}$, corroborating that this
sample is equilibrated at the bath temperature $T=(323\pm30)$K
within the uncertainty of the experiment.

\section{Linear response regime}
\label{linear}

It is important to test for a well-defined linear response regime
for small enough external forces where the mobility becomes
independent of the force. Indeed, previous work \cite{Weeks_force}
did not find a linear regime, but instead there is a threshold
force below which the particles are trapped even under the
influence of the external force. On the contrary, here we find
that if the observable time is larger than the cage life time,
particles are always able to explore the structural relaxation
leading to cage rearrangements, either in the presence or in the
absence of the external force and therefore the linear response
regime ensues. This is because the present experiments focus on
larger time scales than previous work.

\begin{figure}
\centering \resizebox{12cm}{!}{\includegraphics{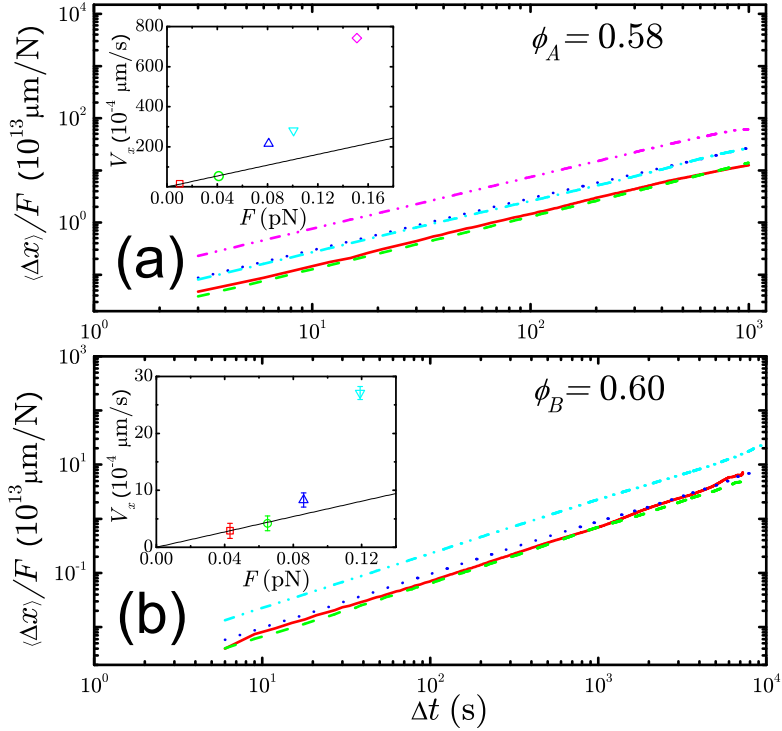}}
\caption{The response functions for (a) sample A and (b) sample B.
The insets are plots of the velocity of magnetic beads as a
function of the force, $V_{x}$ versus $F$. The straight line
indicates the linear response regime. The velocity error is due to
the accuracy to locate the bead positions, smaller than the symbol
size in the inset of (a).} \label{lresponse}
\end{figure}

In order to find the linear response regime, we apply different
forces for samples A and B. The results are shown in Fig.
\ref{lresponse}a and Fig. \ref{lresponse}b. The collapsing of the
results onto a single curve of $\langle\Delta x\rangle/F$ suggest
a linear response for lower forces. The insets show the
relationship between the velocity, $V_x=\langle\Delta
x\rangle/\Delta t$, of the tracers along the force direction and
the force $F$, which indicates that the external force cannot be
larger than $4\times 10^{-14}$N and $6\times 10^{-14}$N to keep
the linear response for samples A and B, respectively. It then
confirms that the force $F=1.7\times 10^{-14}$N used in our
studies is in the linear response regime.

It is important to investigate the age-dependence of the linear
response. As we addressed in the original manuscript, many
measurements have to be taken to reveal the aging behavior for one
value of the external force. Therefore to measure the aging effect
in Fig. \ref{msd}c, we measured over 75 trajectories to achieve a
statistically reliable average. In order to find the region of
forces in the linear response regime, many experiments have to be
run at different forces. Thus, for the determination of the linear
response we are obliged to use a smaller number of tracers
(roughly 10). In this case, in order to improve the statistical
average, we average over the waiting time $t_w$ and obtain a
single velocity $V_x$ for each force, as shown in Fig.
\ref{lresponse}. (The aging dependence is ignored, and the
response functions as shown in Fig. \ref{lresponse} are calculated
by regular time-translation-invariance methods). This average is
only applied to find the value of the external force in the linear
response regime. Once this force is identified, then we perform
the full aging analysis by measuring a larger amount of tracers.


\section{Sedimentation of magnetic tracers}

\begin{figure}
\centering \resizebox{12cm}{!}{\includegraphics{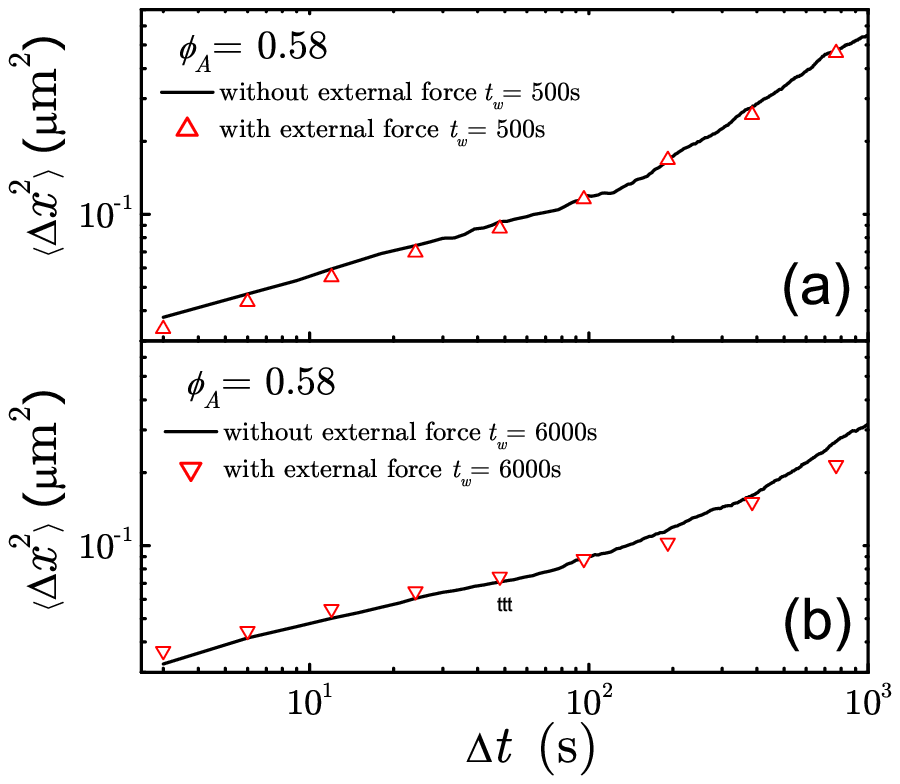}}
\caption{MSD for the sample A at (a) $t_w = 500$s and (b) $t_w =
6000$s analyzed from the data without magnetic force and
with magnetic force, respectively.} \label{sedimentation}
\end{figure}

The density of the magnetic tracers is very close to the background
PMMA particles. Since the sedimentation is proportional to the mismatch
between the density of the tracers and the surrounding liquid
(which is the same as the PMMA density), then we expect that the
effect of sedimentation would be small.

In principle, this effect is negligible in the more dense
sample B since the tracers remain
in the  x-y plane for few hours
without any noticeable sedimentation effect.
In the case of the less concentrated sample A,
sedimentation of tracers can be seen after one hour of observation.
Since we measure up to $t_w=6000$s, it is important to
check the effect of  sedimentation on
the diffusion of particles for this sample.

Unfortunately,
we cannot test directly the effect of this
sedimentation, since we cannot match the
density of the tracers with that of the fluid and PMMA particles
(the tracers are
magnetic and therefore slightly heavier than the PMMA
particles by construction).
However we can indirectly test the main issue at hand:
What is the effect of an
external driving constant force of the order of gravity ($\sim$pN)
on the tracer diffusion?

To this end we  apply a magnetic external force
of the same value as the sedimentation force and measure the
resulting effect on the diffusion constant for sample A.
We found that
the effect of the external force on diffusion is negligible as long
as the force is small enough ($\sim $pN).
Figure \ref{sedimentation} plots   the MSD
for the sample A at (a) $t_w = 500$s and (b) $t_w = 6000$s.
We obtain  the MSD analyzing the trajectories of tracers
without magnetic force and with
magnetic force (in this case, we properly subtract the average value
of the displacements). The figure indicates that
the diffusion is the same with or without the external force,
thus confirming its negligible effect for this case.

\section{Determination of the aging exponents}
\label{exponents}

\begin{figure}
\centering \resizebox{12cm}{!}{\includegraphics{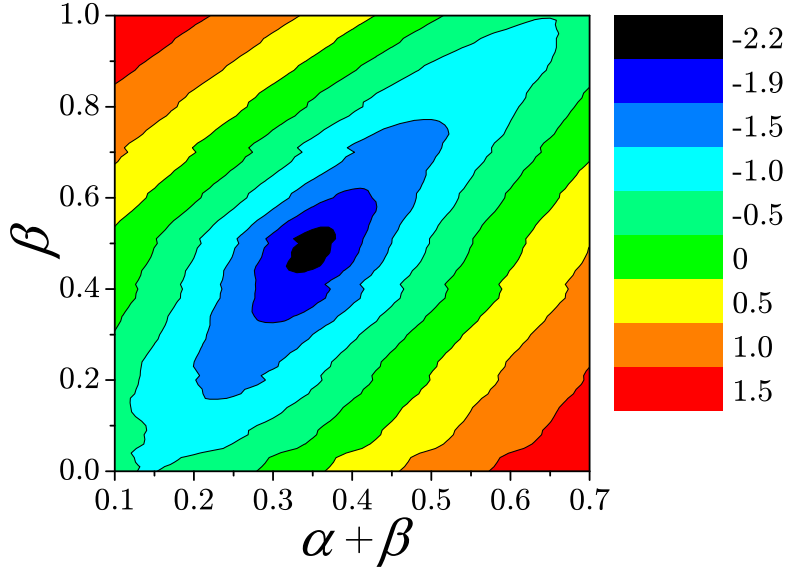}}
\caption{Log plot of standard deviation, $\ln\sigma^2$, which
quantifies the collapsing of $t_w ^{\alpha}C(t_w ,t_w +\Delta t)$
for all the data of sample A (in color scale or gray scale). The
minimum of $\sigma^2$ (black basin) gives the best aging exponents
at $\alpha+\beta=0.34\pm0.05$, $\beta=0.48\pm0.05$.} \label{chi}
\end{figure}

As described in the main manuscript, all the MSD for different
aging times $t_w $ collapse onto a master curve when plotted as
$t_w^{\alpha}C(t_w ,t_w +\Delta t)$ versus $\Delta t/{t_w
}^{\beta}$. In order to search for the best value of the aging
exponents $\alpha$ and $\beta$ that give the correct data
collapse, we calculate the standard deviation of all the curves as
defined below:

\begin{equation}
\begin{split}
&\sigma^2 = \sum_{\ln(\Delta t/t_w^{\beta})}\sum_{\ln(t_w)}\\
&[\ln(t_w^{\alpha}C(t_w,\Delta
t/t_w^{\beta}))-\frac{1}{N}\sum_{\ln(t_w)}\ln(t_w^{\alpha}C(t_w,\Delta
t/t_w^{\beta}))]^2.
\end{split}
\end{equation}

We search a wide range of $\alpha$ and $\beta$ to minimize
$\sigma^2$. The result is plotted in Fig. \ref{chi} showing that
the best data collapsing with $\sigma^2=0.1$ can be achieved at
$\alpha + \beta = 0.34\pm0.05$ and $\beta = 0.48\pm0.05$ as
reported in the main text.

Spin-glass models predict a general scaling form $C_{\rm ag}(t,t_w
) = C_{\rm ag}(h(t)/h(t_w))$, where $h(t)$ is a generic monotonic
function \cite{Cugliandolo}. In particular, a familiar form is
$h(t)\sim t^\mu$ where $\mu$ is a power law exponent giving rise
to the cases of simple aging, $\mu=1$, super aging, $\mu>1$, and
sub-aging, $\mu<1$. Assuming the above form, the linear behavior
$C_{\rm ag}(t_w +\Delta t,t_w ) \sim \Delta t$ for larger $\Delta
t$ requires $\mu=1$. This is just a special case of our scaling
ansatz Eq. (\ref{scaling-laws1}) with $\beta = 1$. As we show in
Fig. \ref{chi}, our data cannot be collapsed for this parameter.

\section{Scaling behavior of local fluctuations}
\label{local}

The local correlation function $C$ and local integrated response
function $\chi$ studied in the manuscript are calculated from each
individual tracer trajectory. In a condensed colloidal sample, the
tracer trajectories are always confined at a local position.
Therefore, the correlation $C$ and response $\chi$ for an
individual particle can be regarded as the coarse-grained {\it
local} fluctuations of the observables as investigated in
\cite{Castillo}.

Furthermore, in order to improve the statistics of our results,
the PDF $P(C)$ is calculated not only for all the tracers, but
also over a time interval much smaller than the age of the system.
In practice, we calculate the $i-$th tracer's local correlation
function as $C_{i}(t_w+\Delta t, t_w)$. At a given $\Delta t$, we
open a small time window $[\Delta t-t_{s},\Delta t+t_{s}]$ ($t_s
\ll \Delta t$) and count all the $C_{i}(t_w+\Delta t_j, t_w)$ with
$\Delta t_j\in[\Delta t-t_{s},\Delta t+t_{s}]$ into the statistics
of $P(C)$. The calculated $P(C)$ is a mixture of the local and the
temporal fluctuations of the observables. Similar technique is
performed to calculate $P(\chi)$.

Below we derive the scaling law for the PDF of the local
correlation function, Let us first recall the scaling behavior of
the global correlation function in Eq. (\ref{scaling-laws}),

\begin{equation}\label{gC}
t_w^{\alpha}\overline{C(t_w +\Delta t,t_w )}=f_D({\Delta t}/{{t_w
}^{\beta}}),
\end{equation}
where we add a bar to $\overline{C}$ to distinguish the global
correlations from the local $C$. The average is taken over all the
tracer particles. We can rewrite the global correlation function
as the integration of the PDF $P(C)$ as:

\begin{equation}\label{lC}
\overline{C}= \int P(C)CdC.
\end{equation}

Furthermore, we obtain the relation of $f_D$ and $P(C)$ by
substituting Eq. (\ref{lC}) into Eq. (\ref{gC}):

\begin{equation}\label{sC}
f_D({\Delta t}/{{t_w }^{\beta}})=t_w^\alpha\overline{C}= \int
t_w^{-\alpha}P(C)(t_w^\alpha C) d(t_w^\alpha C).
\end{equation}
For a given ${\Delta t}/{{t_w }^{\beta}}$, $f_D$ is equal to a
constant, and Eq. (\ref{sC}) requires that $t_w^{-\alpha}P(C)$
should only depend on $t_w^\alpha C$. In other words,
$t_w^{-\alpha}P(C)$ is a function of ${\Delta t}/{{t_w }^{\beta}}$
and $t_w^\alpha C$. Then we define $F_D$ as:

\begin{equation}
F_D(t_w^\alpha C, \Delta t/t_w^\beta)=t_w^{-\alpha}P(C).
\end{equation}
A similar formula can be obtained for $P(\chi)$. Eventually, we
obtain the scaling ansatz of $P(C)$ and $P(\chi)$ shown in Fig.
\ref{pdf}:
\begin{subequations}
\begin{align}
P(C)&=t_w^{\alpha}F_D(t_w^\alpha C, \Delta
t/t_w^\beta),\\
P(\chi)&=t_w^{\alpha}F_M(t_w^\alpha \chi, \Delta t/t_w^\beta),
\end{align}
\end{subequations}
where the universal functions $F_D(x,y)$ and $F_M(x,y)$ satisfy
\begin{subequations}
\begin{align}
\int F_D(x,y)dx = f_D(y),\\
\int F_M(x,y)dx = f_M(y).
\end{align}
\end{subequations}

\section{Study of the PDF of the autocorrelation function}
\label{power law}

The PDF of the autocorrelation function follows a modified power
law $t_w^{-\alpha}P(C)\propto(t_w^\alpha C+C_0)^{-\lambda}$, as we
see in Figs. \ref{pdf}a, \ref{pdf}b and \ref{pdf}c. From Eq.
(\ref{sC}), we have
\begin{equation}
\begin{split}
t_w^\alpha\overline{C}&= \int_{0}^{C_{cut}}
t_w^{-\alpha}P(C)(t_w^\alpha C) d(t_w^\alpha C)\\
&= \frac{\int_{0}^{C_{cut}} (x+C_0)^{-\lambda} x
dx}{\int_{0}^{C_{cut}} (x+C_0)^{-\lambda} dx}.
\end{split}
\end{equation}
The cutoff $C_{cut}$ ($C_{cut}\gg C_0$) is introduced to make the
integral converge and we always take $\lambda \geq 1$, then
\begin{equation}
t_w^\alpha\overline{C}=
\frac{C_0}{\lambda-2}[1-(\lambda-1)(\frac{C_{cut}}{C_0}+1)^{2-\lambda}].
\label {C0Ccut}\end{equation}

For $\lambda > 2$, the last term in Eq. (\ref{C0Ccut}) is
negligible and $t_w^\alpha\overline{C}\approx C_0/(\lambda-2)$
mainly depends on the short-time parameter $C_0$. For $\lambda < 2$,
we have $t_w^\alpha\overline{C}\approx
\frac{\lambda-1}{2-\lambda}C_0(\frac{C_{cut}}{C_0})^{2-\lambda}$
and the long-time parameter $C_{cut}$ dominates.

Following the previous discussion of $P(C)$, we define the $i$-th
tracer's local correlation function as:

\begin{equation}
C_{i}(t_w+\Delta t, t_w)=\langle\Delta x(t_w+\Delta
t,t_w)^2\rangle_i/2,
\end{equation}
where the average $\langle ... \rangle_i$ is calculated for only
the $i$-th tracer's trajectory by opening a small time window
$[\Delta t-t_{s},\Delta t+t_{s}]$ ($t_s \ll \Delta t$) at a given
$\Delta t$, and counting all the $C_{i}(t_w+\Delta t_j, t_w)$ with
$\Delta t_j\in[\Delta t-t_{s},\Delta t+t_{s}]$ into the statistics
of $P(C)$. We should note that if there is no average $\langle ...
\rangle_i$, $P(C)$ can be reduced to a simple form:

\begin{equation}
P(C)=P(\langle\Delta x^2\rangle_i/2)\xrightarrow{\langle ...
\rangle_i\rightarrow 0} P(\Delta x^2/2).
\end{equation}
This form can be further reduced since
\begin{equation}
P(\Delta x^2/2)d(\Delta x^2/2)=P(\Delta x)dx,
\end{equation}
therefore at the limit of the average $\langle ...
\rangle_i\rightarrow 0$,
\begin{equation}\label{Px}
P(C)=P(\langle\Delta x^2\rangle_i/2)\xrightarrow{\langle ...
\rangle_i\rightarrow 0} P(\Delta x^2/2)=P(\Delta x)/{\Delta x}.
\end{equation}

Therefore our $P(C)$ is relate to the probability $P(\Delta x)$
usually studied in previous works \cite{Weeks_sci}. We find that
$P(\Delta x)$ can be approximately fitted by a broad tail power
law behavior for large $\Delta x$, consistent with equation
\ref{Px}. However the data of $P(\Delta x)$ can be better fitted
by stretch exponential rather than power law, which is consistent
with previous works \cite{Weeks_sci}. On the contrary,
$P(\langle\Delta x^2\rangle_i/2)$ can not be fitted by stretch
exponential. Thus we believe that the power law fit is more proper
to describe our result of $P(C)$. The difference with other
studies might be due to the fact that our samples
are glassy while others work in
the supercooled regime.





\end{document}